# Oncolytic mechanisms and immunotherapeutic potential of Newcastle disease virus in cancer therapy

Umar Ahmad[1,2]*, Surializa Harun[1], Moussa Moise Diagne[3], Syahril Abdullah[4], Khatijah Yusoff [4], Abhi Veerakumarasivam[4,5]

[1]Human Genetics Informatics, Department of Anatomy, Faculty of Basic Medical Sciences, Sa'adu Zungur University, Bauchi State, PMB 65, Itas/Gadau, Nigeria. [2]Institute of Genomics, Centre for Laboratory Diagnostics and Systems, Africa Centres for Disease Control and Prevention, P.O. Box 200050, Addis Ababa, Ethiopia. [3]Virology Department, Institut Pasteur de Dakar, Dakar 220, Senegal. [4]Malaysia Genome and Vaccine Institute, National Institutes of Biotechnology Malaysia, Jalan Bangi, 43000 Kajang, Selangor Darul Ehsan, Malaysia.[5]Department of Biological Sciences, School of Medical and Life Sciences, Sunway University, 47500 Bandar Sunway, Selangor Darul Ehsan, Malaysia.

*Corresponding author: umarahmad@basug.edu.ng

**Abstract**

Newcastle Disease Virus (NDV), classified as *Avian orthoavulavirus 1* (avian paramyxovirus type-1), is a promising oncolytic agent that selectively targets and destroys cancer cells while sparing normal tissues. Its oncoselectivity exploits cancer-specific defects in antiviral defenses, particularly impaired Type I interferon signaling, and dysregulated apoptotic pathways, enabling robust viral replication and cytotoxicity in malignancies such as breast, colorectal, and melanoma. NDV induces intrinsic and extrinsic apoptosis through caspase activation and triggers immunogenic cell death via damage-associated molecular patterns, stimulating potent antitumours immune responses. Additionally, NDV's potential as a vaccine vector, expressing tumours-associated antigens, offers prospects for prophylactic and therapeutic cancer applications. This review provides a comprehensive analysis of NDV's morphology, classification, and molecular biology, focusing on its viral entry and replication mechanisms in host cells. It explores NDV's interactions with cancer cells, emphasizing its ability to induce cytotoxicity and immune activation. Understanding these mechanisms is critical for optimizing NDV's oncolytic potential and advancing its clinical translation. Future directions include enhancing NDV through genetic engineering, combining it with therapies like immune checkpoint inhibitors, and developing personalized medicine approaches tailored to tumours genomic profiles. These advancements position NDV as a versatile therapeutic agent in oncolytic virotherapy.

## 1.0 Introduction

Newcastle disease virus (NDV), an avian paramyxovirus type-1 (APMV-1), belongs to the *Avulavirus* genus, within the subfamily *Paramyxovirinae*, family *Paramyxoviridae*, and order *Mononegavirales* [1-5]. Morphologically, NDV is pleomorphic, with a spherical structure that ranges from 100 to 300 nm in diameter. Its lipid bilayer envelope, derived from the host cell membrane, is relatively unstable and houses critical viral proteins [6-8]. NDV is a pathogen of significant economic importance, primarily affecting poultry, particularly chickens– resulting in high mortality, decreased egg production, and increased costs of vaccination and biosecurity[9]. Transmission occurs via inhalation, and the clinical severity in birds is determined by the viral pathotype [8].

NDV strains are classified into three pathotypes—velogenic, mesogenic, and lentogenic—based on their virulence and clinical impact, as measured by the Intracerebral Pathogenicity Index (ICPI) and Mean Death Time (MDT) [1, 5]. Velogenic strains cause severe intestinal and neurological disease with high mortality, mesogenic strains result in moderate respiratory and nervous system disease, and lentogenic strains typically cause mild respiratory symptoms [1, 10]. Notably, NDV's oncolytic potential correlates with its pathotypic classification; velogenic strains exhibit robust replication in human cancer cells, leading to efficient cell lysis, whereas lentogenic strains are less virulent due to restricted F0 protein cleavage [10, 11]. These distinctions in virulence, oncolytic activity, and oncoselectivity are summarized in **Table 1**, which highlights the ICPI and MDT ranges for each pathotype [5, 7, 10, 12-14].

The inherent pathogenic diversity of NDV, driven by its classification into lytic (velogenic and mesogenic) and non-lytic (lentogenic) strains, underpins its promise as an oncolytic agent for cancer therapy [10, 14-20]. Velogenic strains, such as AF2240 and PV701, exhibit robust replication and cytotoxicity in human cancer cells, including breast, colorectal, and melanoma, due to efficient F0 protein cleavage, while lentogenic strains like LaSota and Hitchner B1 offer selective tumour lysis with lower virulence. Beyond direct oncolysis, NDV's ability to induce immunogenic cell death (ICD) and cytokine production positions it as a versatile platform for cancer immunotherapy, with strains engineered to express tumours-associated antigens enhancing both prophylactic and therapeutic immune responses [21, 22]. These dual roles highlight NDV's potential to target tumourss while stimulating systemic antitumours immunity, broadening its therapeutic applications.

This review synthesizes the current understanding of NDV's oncolytic mechanisms, emphasizing its selective targeting of cancer cells through defects in antiviral defenses, apoptotic pathways, and growth signaling, as well as its induction of ICD. By exploring NDV's molecular biology, viral entry, and replication, alongside its cytotoxic and immunotherapeutic effects, we aim to elucidate pathways critical for optimizing its clinical translation in cancer therapy, with future directions including genetic engineering and combination therapies [23, 24].

**2.0 Newcastle disease virus**

Newcastle disease (ND), a highly contagious avian illness, was first identified in 1926 during an outbreaks in Newcastle-on-Tyne, England, and the island of Java (now Indonesia) [7]. While some reports suggest earlier outbreaks, these claims remain speculative [4]. The causative agent, Newcastle disease virus (NDV), also known as avian paramyxovirus type-1 (APMV-1), was later characterised and formally classified by Alexander and Allan in 1974 [1]. This virus is economically significant, as it can cause considerable losses in the poultry industry [8]. NDV primarily affects chickens, but it has a broad host range, infecting 27 of the 50 bird orders [13, 25].

## 2.1 Morphology and characteristics

NDV exhibits an unstable, pleomorphic structure, capable of changing shape and size in response to external stimuli or environmental factors [6, 7, 26]. The virus is enveloped by a lipid bilayer derived from the host cell membrane, within which two glycoproteins, haemagglutination-neuraminidase (HN) and fusion (F) proteins, are embedded [8]. Beneath this lipid membrane, matrix proteins are attached to the inner surface, providing structural support. The viral genome consists of non-segmented, single-stranded, negative-sense RNA, which is encapsulated by nucleocapsid (NP) proteins. The nucleocapsid associates with phosphoproteins (P) and large polymerase (L) proteins, forming a ribonucleoprotein complex essential for viral replication [8]. This virion structure is illustrated in **Figure 1**, which highlights the arrangement of NP, P, M, F, HN, and L proteins within the NDV virion [11].

## 2.2 Classification of NDV

NDV also known as avian paramyxovirus type-1 (APMV-1), is classified as *Avian orthoavulavirus 1* within the genus *Orthoavulavirus*, subfamily *Avulavirinae*, and family *Paramyxoviridae* [26, 27]. This family includes notable viruses such as mumps, human parainfluenza, Sendai virus, simian virus-5, and the more recently identified Nipah and Hendra viruses[27]. . In 1993, the International Committee on the Taxonomy of Viruses reorganized the paramyxovirus genus and assigned NDV to the *Rubulavirus* genus within the *Paramyxovirinae* subfamily [12]. However, due to NDV's lack of the small hydrophobic (SH) gene characteristic of rubulaviruses, it was reclassified into the *Avulavirus* genus and subsequently the *Orthoavulavirus* genus to reflect its genetic and phylogenetic distinctions [27, 28].

## 2.3 Molecular biology of NDV and its functional proteins

The genome of NDV, a non-segmented, negative-sense, single-stranded RNA of approximately 15 to 16 kb, encodes six essential genes producing six structural proteins in the sequence 3'-NP-P-M-F-HN-L-5', along with two non-structural proteins, V and W (Dimitrov et al., 2019; [5, 7]). This genome organization is depicted in **Figure 2**, which illustrates the arrangement of genes encoding the nucleoprotein (NP, 55 kDa), phosphoprotein (P, 53 kDa), and large polymerase protein (L, 200 kDa), which form the nucleocapsid, critical for viral replication [10, 15, 25]. The haemagglutinin-neuraminidase (HN, 74 kDa) and fusion glycoproteins (F, 67 kDa) are embedded in the viral lipid bilayer, facilitating entry, while the matrix protein (M, 40 kDa) supports virion assembly and budding [10, 15, 25]. The functions and oncolytic roles of these proteins, detailed in **Table 2**, highlight their contributions to NDV's tumours-selective infection and cytotoxicity [1, 4, 10, 11, 14, 16-19].

The HN protein mediates the binding of the virus to cell surface receptors, facilitating viral attachment through its haemagglutination and neuraminidase activities [16]. The F protein is responsible for mediating the fusion between the viral envelope and the host cell membrane [14, 18], allowing viral entry. The cleavage of the F protein is a major determinant of NDV virulence. Synthesized as an inactive precursor (F0), it is activated by cellular proteases [20]. In velogenic strains, F0 has polybasic cleavage sites, which are cleaved by furin proteases, enabling systemic infection [12, 13]. In contrast, lentogenic strains contain monobasic cleavage sites that are cleaved by trypsin-like proteases, restricting the infection to digestive and respiratory tracts. Infected cells express highly fusogenic F proteins on their surface, allowing the formation of syncytia, which facilitates viral spread between cells [29-31]. The co-expression of F and HN proteins leads to cell fusion and is critical for viral infectivity and virulence [32, 33].

The V and W non-structural proteins are generated via mRNA editing during P gene transcription [34]. The V protein plays a crucial role in inhibiting interferon responses [35, 36] and inducing apoptosis in chicken cells, but not in human cells, making it a determinant of host range specificity [37, 38]. The W protein also contributes to viral replication and pathogenesis [37].

## 2.4 Viral entry and replication cycle

NDV exhibits a remarkable ability to infect a diverse array of human cancer cells in vitro, including those derived from breast, brain, cervical (HeLa), colorectal, and bladder cancers [1, 39]. These processes, driven by the Haemagglutinin-Neuraminidase (HN) and Fusion (F) glycoproteins, orchestrate attachment, membrane fusion, and genome replication, as illustrated in **Figure 3**, which depicts the complete NDV infectious cycle from receptor binding to virion release [11]. The HN protein mediates attachment to sialic acid receptors, while the F protein facilitates fusion, enabling the viral genome's entry into the cytoplasm for replication [40].

### 2.4.1 Viral attachment

NDV infection begins with the HN protein binding to sialic acid-containing receptors on the host cell surface, primarily α2,6- or α2,3-linked sialic acids, depending on the viral strain [2, 40]. These receptors are ubiquitous across normal and cancerous cells, but their overexpression in tumours cells, particularly α2,6-linked sialic acids in human cancers, enhances NDV's oncolytic selectivity [39]. The HN protein's receptor-binding domain, located in its globular head, rapidly mediates attachment within minutes, enabling NDV to adhere to target cells and erythrocytes, a process critical for initiating infection [2, 11]. This binding triggers conformational changes in HN, exposing a site that interacts with the F protein to coordinate the subsequent fusion step [41].

**2.4.2 Viral fusion and entry**

The F protein facilitates fusion of the viral envelope with the host cell membrane, a process essential for delivering the viral genome into the cytoplasm. Synthesized as an inactive precursor (F0), the F protein requires cleavage by host proteases into its active, disulfide-linked F1 and F2 subunits [11, 39]. The cleavage site's structure is a key determinant of NDV virulence: velogenic strains possess polybasic cleavage sites, cleaved by ubiquitous furin-like proteases, enabling systemic infection, while lentogenic strains have monobasic sites, cleaved by trypsin-like proteases, restricting infection to specific tissues [40]. Upon HN binding to sialic acid receptors, the HN-F interaction induces a conformational shift in the F protein, exposing its hydrophobic fusion peptide, which inserts into the host membrane to initiate fusion [41-43]. Electron microscopy reveals two primary entry pathways: direct fusion at the plasma membrane, where the viral envelope merges with the cell membrane to release the viral nucleocapsid, and endocytosis, where NDV is internalized into endocytic vesicles, and the viral envelope fuses with the vesicle membrane to release the nucleocapsid into the cytoplasm [19, 44, 45]. The choice of pathway depends on host cell type and protease availability, with direct fusion being more efficient in tumours cells with high surface protease activity [11].

**2.4.3 Replication cycle**

Once the viral nucleocapsid, comprising the negative-sense RNA genome, nucleoprotein (NP), phosphoprotein (P), and large polymerase (L) proteins, enters the cytoplasm, NDV initiates replication without nuclear involvement, a hallmark of *Paramyxoviridae* [26, 28, 42, 43]. The viral RNA-dependent RNA polymerase, formed by the P and L proteins, transcribes the genome into monocistronic mRNAs for viral protein synthesis, driven by L's methyltransferase activity for mRNA capping [46]. The NP protein encapsidates the RNA, protecting it from host nucleases and facilitating polymerase access [47-49]. Initially, transcription predominates, producing mRNAs for HN, F, M, NP, P, and L proteins. As NP accumulates, it triggers a switch to replication, where the polymerase synthesizes full-length, positive-sense antigenomes, which serve as templates for negative-sense genomic RNA [26, 42, 43]. The P protein's V variant, produced via RNA editing, further enhances replication by degrading STAT1, suppressing host interferon responses, a critical factor in NDV's tumours-selective replication [5].

### 2.4.4 Viral assembly and release

Newly synthesized viral proteins and genomes assemble in the cytoplasm, coordinated by the Matrix (M) protein, which interacts with the RNP complex and envelope glycoproteins (HN and F) at the host cell membrane [49]. The HN protein's neuraminidase activity cleaves sialic acid residues, preventing viral aggregation and facilitating virion release through budding [39]. This efficient replication and release cycle sustains high viral titers, particularly in tumours cells with defective antiviral defenses, amplifying NDV's oncolytic effects [11].

### 2.4.5 Oncolytic implications

NDV's entry and replication mechanisms underpin its oncolytic potency. The overexpression of sialic acid receptors on cancer cells enhances viral attachment, while the F protein's polybasic cleavage in virulent strains promotes efficient fusion and syncytium formation, inducing apoptosis and immunogenic cell death [39]. The replication cycle's reliance on cytoplasmic machinery and suppression of interferon responses via the V protein enables selective replication in tumours cells, sparing normal cells with intact antiviral pathways [5, 50]. These features make NDV a promising oncolytic agent, capable of triggering robust antitumours immune responses through cytokine induction and tumours antigen exposure.

## 3.0 Newcastle disease virus and Cancer

NDV is a potent oncolytic agent, capable of stimulating host immune responses and inducing cytokine production in tumuor cells, thereby promoting robust anticancer activity [2, 39]. Naturally occurring NDV strains have demonstrated efficacy in preclinical models, including various animal tumours models and cancer cell lines such as breast, colorectal, and melanoma [16, 23, 39, 44]. Clinical studies have explored NDV's potential as an oncolytic vaccine in humans, with trials targeting malignant melanoma [19, 50, 51], head and neck squamous cell carcinomas [52-54], colorectal carcinoma [55, 56], digestive tract tumourss [53, 57]), and other advanced cancers. A range of NDV strains, including mesogenic (73-T, Beaudette C), velogenic (Italian, PV701, HUJ, AF2240), and lentogenic (MTH68, Ulster, V4-UPM, Hitchner B1, LaSota) strains, as well as others like S, Ijuk, F, and 0I/C, have shown significant oncolytic activity across these models, summarized in **Table 3** [24, 58-61]. Notably, Malaysian strains AF2240 (velogenic) and V4-UPM (lentogenic) have exhibited promising results in cancer research, with studies by Meyyappan [62] demonstrating their efficacy against human breast cancer cell lines, highlighting their potential as effective anticancer agents [8, 63]. The inclusion of widely studied strains like Hitchner B1 and LaSota, known for selective tumours cell lysis, and Beaudette C, effective in glioma and melanoma, further underscores NDV's versatility in oncolytic therapy [11, 24].

## 4.0 Mechanism of oncoselectivity and cytotoxic

### 4.1 Defects in antiviral pathways

NDV exploits inherent defects in the interferon (IFN) signaling pathways of cancer cells, leading to enhanced viral replication in malignant tissues compared to normal cells, a key factor in its selective oncolytic activity [1, 2, 39]. Type I interferons (IFN-α/β) are critical for host antiviral defense, activating the JAK-STAT signaling pathway to induce interferon-stimulated genes (ISGs) that inhibit viral replication and promote apoptosis in infected cells [2, 64]. However, cancer cells frequently exhibit impaired IFN signaling due to mutations or epigenetic silencing of JAK-STAT components, resulting in a diminished antiviral state [16, 65, 66]. This defective response allows NDV to replicate efficiently in tumours cells, while normal cells with intact IFN pathways rapidly upregulate ISGs to restrict viral spread [39]. The NDV V protein, produced via RNA editing of the P gene, further enhances oncoselectivity by degrading STAT1, suppressing IFN responses in cancer cells, and promoting sustained viral replication [5]. The robust replication in tumours cells triggers cytotoxic effects, including apoptosis, necrosis, and immunogenic cell death, driven by syncytium formation (mediated by the F protein) and cytokine release, which amplify antitumours immunity [11]. These mechanisms underpin NDV's oncoselectivity and cytotoxicity, making it a promising candidate for oncolytic virotherapy [23].

### 4.2 Defects in apoptotic pathways

Cancer cells often exhibit dysregulated apoptotic pathways, which NDV leverages by activating both intrinsic and extrinsic apoptosis pathways, resulting in effective cell death [19, 44]. NDV exploits these defects to induce potent cytotoxic effects, with the activation of caspase cascades further underlining its efficacy [39]. In the intrinsic pathway, NDV triggers mitochondrial outer membrane permeabilization (MOMP) through upregulation of pro-apoptotic Bcl-2 family proteins (e.g., Bax, Bak), leading to cytochrome c release and activation of caspase-9, which initiates the caspase cascade involving executioner caspases like caspase-3 [34]. The extrinsic pathway is activated by NDV-induced production of tumours necrosis factor-alpha (TNF-α) and TNF-related apoptosis-inducing ligand (TRAIL), which bind to death receptors (e.g., TNFR1, DR4/DR5), recruiting caspase-8 to form the death-inducing signaling complex (DISC) and amplifying caspase activation [19, 44, 46]. Cancer cells frequently overexpress anti-apoptotic proteins (e.g., Bcl-2, XIAP) or have impaired pro-apoptotic signaling, which ND umber of NDV strains, such as AF2240 and V4-UPM, have shown enhanced apoptosis in breast cancer cell lines despite these defects, indicating NDV's ability to overcome resistance to apoptosis [50, 62, 67]. Notably, NDV-induced apoptosis can occur independently of functional IFN signaling, as the virus exploits synergistic defects in IFN and apoptotic pathways, enhancing its oncoselectivity and cytotoxicity in tumours cells with defective antiviral defenses [23]. These mechanisms make NDV a promising oncolytic agent, capable of inducing immunogenic cell death and robust antitumours immunity [11].

**4.3 Other pathways of oncoselectivity**

NDV enhances its oncoselectivity through a multifaceted approach, critical for harnessing its full potential as an effective therapeutic agent in cancer treatment [39]. A key mechanism is the induction of immunogenic cell death (ICD) in cancer cells, where NDV infection triggers the release of damage-associated molecular patterns (DAMPs) such as calreticulin, ATP, and high-mobility group box 1 (HMGB1). These DAMPs act as "danger signals," promoting dendritic cell maturation, antigen presentation, and activation of cytotoxic T cells, which amplify the host's immune response against infected tumours cells [39, 68]. This immune-mediated cytotoxicity significantly enhances NDV's oncolytic effects, fostering a robust antitumours immune response [23]. Additionally, NDV preferentially replicates in cancer cells with dysregulated growth signaling pathways, such as overactive RAS, PI3K/AKT, or EGFR pathways, which are common in tumourss and impair antiviral defenses, further promoting viral replication and cytotoxicity [43, 69, 70]. These pathways often upregulate anti-apoptotic signals, which NDV counteracts through its pro-apoptotic mechanisms, ensuring selective tumours cell death [11].

In summary, NDV's selective infection and killing of cancer cells result from a synergistic combination of defects in antiviral pathways (e.g., impaired JAK-STAT signaling), dysregulation of intrinsic and extrinsic apoptotic pathways, induction of immunogenic cell death (ICD) via DAMPs, and preferential replication in tumours cells with dysregulated growth signaling (e.g., RAS, PI3K/AKT). These mechanisms, detailed in **Table 4**, underscore NDV's multifaceted approach as a potent therapeutic agent in oncolytic virotherapy, capable of targeting tumourss while stimulating systemic antitumours immunity [5, 11, 19, 23, 24, 34, 39, 41, 44, 50, 68, 71-74]

**5.0 Conclusion and future perspective**

NDV classified as *Avian orthoavulavirus 1*, represents a highly promising oncolytic agent due to its ability to selectively target and eliminate cancer cells while sparing normal tissues, offering a transformative approach to cancer therapy [39]. Its oncoselectivity is driven by exploiting cancer-specific vulnerabilities, including defective Type I interferon (IFN) signaling via the JAK-STAT pathway, which allows robust viral replication in tumours cells with impaired interferon-stimulated gene (ISG) expression [39, 72]. NDV further capitalizes on dysregulated apoptotic pathways, inducing intrinsic apoptosis through mitochondrial outer membrane permeabilization (MOMP) and caspase-9 activation, and extrinsic apoptosis via TNF-α and TRAIL-mediated caspase-8 activation, overcoming anti-apoptotic barriers like Bcl-2 overexpression [44]. The induction of immunogenic cell death (ICD), marked by the release of damage-associated molecular patterns (DAMPs) such as calreticulin, ATP, and HMGB1, stimulates dendritic cell maturation and cytotoxic T-cell activation, fostering systemic antitumours immunity [68]. Additionally, NDV's preferential replication in tumours cells with dysregulated growth signaling pathways, such as RAS or PI3K/AKT, enhances its cytotoxic potency [70]. These multifaceted mechanisms, exemplified by strains like AF2240, PV701, and LaSota, underpin NDV's efficacy across diverse cancers, including breast, colorectal, and melanoma, as demonstrated in preclinical and clinical studies [24, 54, 75-77].

Despite these advances, several challenges must be addressed to fully integrate NDV into mainstream cancer therapy. Optimizing delivery methods, such as developing nanoparticle-based carriers or intratumoursal injections to enhance tumours penetration and bioavailability, is critical, particularly for solid tumours with dense stromal barriers [11]. Improving tumour specificity requires addressing variability in sialic acid receptor expression (e.g., α2,6- vs. α2,3-linked) and overcoming resistance mechanisms, such as upregulated anti-apoptotic proteins (e.g., XIAP) or restored IFN signaling in resistant tumours subsets [39, 40]. Immune clearance by neutralizing antibodies and rapid viral clearance in the bloodstream further limit systemic efficacy, necessitating strategies to enhance NDV's stability and immune evasion [23]. Genetic engineering offers promising solutions, with studies demonstrating enhanced oncolytic activity through modifications like incorporating therapeutic transgenes (e.g., IL-2, GM-CSF) or altering the F protein's polybasic cleavage site to increase tumours-specific fusion efficiency [39].

Future research should prioritize synergistic combination therapies to amplify NDV's therapeutic impact. Pairing NDV with immune checkpoint inhibitors (e.g., anti-PD-1/PD-L1) can enhance T-cell-mediated tumours clearance, while combining with chemotherapy or targeted therapies (e.g., EGFR inhibitors) may overcome tumours microenvironment resistance, as seen in preclinical models of colorectal and glioma cancers [24]. Personalized medicine approaches, leveraging genomic profiling to match NDV strains to tumours-specific mutations (e.g., RAS or EGFR dysregulation), hold significant potential for tailoring therapies to individual patients, as demonstrated with strains like Beaudette C in glioma [70]. Emerging technologies, such as CRISPR-based editing to optimize NDV's HN or F proteins or nanoparticle-mediated delivery to improve tumours targeting, could further enhance specificity and efficacy [11]. Additionally, exploring NDV as a vaccine vector, expressing tumours-associated antigens to elicit robust humoral and cellular immunity, aligns with recent advances in cancer immunotherapy [77, 78]. Continued investigation into NDV's molecular interactions with tumour signaling pathways and immune modulation will be crucial to overcome these challenges and translate its potential into clinical practice.

NDV's unique ability to combine direct oncolysis with immune stimulation positions it as a cornerstone for innovative cancer therapies. With ongoing advancements in virology, bioengineering, and personalized immunotherapy, NDV has the potential to revolutionize cancer treatment, offering hope for improved outcomes across a wide range of ma

**Acknowledgement**

**Funding**

This review received no specific funding support from any public, commercial, or not-for-profit agencies.

**Author contributions**

UA conceptualized the study, wrote the manuscript, and coordinated the review process, while SH designed the NDV figures. SA conducted literature reviews and assisted in drafting the molecular biology section. MMD provided expertise on NDV classification and replication mechanisms and revised the manuscript. KY supervised the project, provided critical insights on NDV's oncolytic mechanisms, and revised the manuscript. All authors read and approved the final manuscript.

**Competing interests**

The author declares that they have no competing interests.

**Table 1: Comparison of NDV pathotypes.**

| Pathotype | Virulence | Oncolytic Activity | Oncoselectivity | ICPI Range | MDT (Hours) | References |
|---|---|---|---|---|---|---|
| Velogenic | High | Strong | High | 1.5–2.0 | 48–72 | [5, 12, 46] |
| Mesogenic | Moderate | Moderate | Moderate | 0.7–1.5 | 72–120 | [7, 13] |
| Lentogenic | Low | Weak | Low | 0.0–0.7 | >120 | [10, 14, 49] |

**Table 2: Key features of NDV functional proteins**

| Protein | Molecular weight (kDa) | Function | Role in oncolysis | References |
|---|---|---|---|---|
| Haemagglutinin-Neuraminidase (HN) | 74 | Recognizes and binds to sialic acid-containing receptors (preferentially α2,6- or α2,3-linked, depending on strain) on host cells, facilitating viral attachment; cleaves sialic acid via neuraminidase activity to promote viral release; mediates hemagglutination to enhance viral spread. | Targets tumour cells overexpressing sialic acid receptors (e.g., α2,6-linked in certain cancers); promotes efficient viral entry, initiating oncolytic infection; enhances immune stimulation through antigen presentation and cytokine induction. | [14, 16, 39] |
| Fusion protein (F) | 67 | Mediates membrane fusion via cleavage of F0 into F1 and F2 subunits at a strain-specific cleavage site (polybasic in velogenic strains, monobasic in lentogenic); promotes viral entry and cell-to-cell spread through syncytium formation. | Enhances tumour cell lysis via polybasic cleavage site in virulent strains, increasing fusion efficiency; induces syncytia, triggering apoptosis and immunogenic tumour cell death; facilitates viral spread within tumour masses. | [11, 18, 19] |
| Matrix protein (M) | 40 | Coordinates virion assembly by interacting with HN, F, and nucleocapsid; drives budding of infectious particles from host cell membranes; stabilizes virion structure during release. | Ensures efficient viral replication and release in tumour cells, sustaining high viral titers; supports oncolytic activity by maintaining robust viral propagation within the tumour microenvironment. | [17, 49] |
| Nucleoprotein (NP) | 55 | Encapsulates viral RNA to form the ribonucleoprotein (RNP) complex, protecting the genome from host nucleases; interacts with P and L proteins to regulate replication and transcription. | Maintains genome integrity in tumour cells, enabling sustained viral replication; supports high viral loads, amplifying oncolytic effects and immunogenic cell death in tumours. | [10, 26] |
| Phosphoprotein (P) | 53 | Acts as a cofactor in the polymerase complex, linking NP and L proteins; regulates RNA synthesis; produces V and W proteins via RNA editing, with | Promotes tumour-selective replication by V protein-mediated suppression of interferon responses in cancer cells; enhances viral | [4, 5, 10] |

| Protein | Molecular weight (kDa) | Function | Role in oncolysis | References |
|---|---|---|---|---|
| | | V protein degrading STAT1 to inhibit interferon signalling, modulating host immunity. | persistence, amplifying oncolytic activity and immune-mediated tumour clearance. | |
| Large polymerase (L) | 200 | Catalyses RNA synthesis, including genome replication and mRNA transcription; performs mRNA capping and methylation via methyltransferase activity, ensuring efficient viral gene expression. | Drives robust replication in tumour cells with defective antiviral defences (e.g., impaired interferon pathways); sustains high viral gene expression, enhancing immunogenic cell death and tumour destruction. | [1, 4, 46] |

### Table 3: NDV strains and oncolytic activity

| NDV Strain | Pathotype | Cancer types tested | Oncolytic activity | References |
|---|---|---|---|---|
| 73-T | Mesogenic | Colorectal, Melanoma, Pancreatic | High: Induces significant tumour regression and syncytium formation; stimulates robust antitumour immune responses. | [39, 55, 67] |
| MTH68 | Lentogenic | Head and Neck, Colorectal, Lung | Moderate: Causes partial tumour cell lysis with limited syncytia; induces modest immune activation. | [23, 55, 79] |
| PV701 (MK107) | Velogenic | Breast, Brain, Bladder, Ovarian | High: Promotes extensive apoptosis and syncytium formation; triggers strong cytokine-mediated immune responses. | [11, 54, 80, 81] |
| AF2240 | Velogenic | Breast, Colorectal Leukemia | High: Induces rapid tumour cell death via apoptosis and necrosis; enhances antitumour immunity through cytokine release. | [39, 75, 76, 82] |
| V4-UPM | Lentogenic | Breast, Colorectal, Cervical (HeLa) | Moderate: Achieves selective tumour cell infection with limited cytotoxicity; supports mild immune stimulation. | [23, 24, 83, 84] |
| Hitchner B1 | Lentogenic | Breast, Lung, Prostate | Moderate: Causes selective tumour cell lysis with minimal syncytia; promotes moderate immune activation via tumour antigen exposure. | [23, 24] |
| LaSota | Lentogenic | Colorectal, Pancreatic, Ovarian | Moderate: Induces tumour cell apoptosis with low cytotoxicity; enhances antitumour immunity through interferon induction. | [24, 39] |
| Beaudette C | Mesogenic | Brain (Glioma), Melanoma, Lung | High: Promotes significant tumour regression through syncytium formation and apoptosis; stimulates strong immune responses. | [11, 24] |

**Table 4: Mechanisms of NDV-Mediated Oncolysis**

| Mechanism | Pathway involved | Effect on cancer cells | References |
|---|---|---|---|
| Defects in antiviral response | Impaired Type I IFN signalling (JAK-STAT pathway); NDV V protein-mediated STAT1 degradation | Enables robust viral replication due to reduced interferon-stimulated gene (ISG) expression; promotes tumour-selective infection and amplification of cytotoxic effects in cells with defective antiviral defences. | [5, 39, 71, 72] |
| Intrinsic apoptosis pathway | Mitochondrial outer membrane permeabilization (MOMP) via Bax/Bak; cytochrome c release; caspase-9 and caspase-3 activation | Induces mitochondrial disruption, leading to apoptosis and cell death; overcomes anti-apoptotic Bcl-2 overexpression in cancer cells, enhancing tumour cell lysis and immunogenic cell death. | [11, 34, 44, 50] |
| Extrinsic apoptosis pathway | TNF-α and TRAIL binding to death receptors (TNFR1, DR4/DR5); formation of death-inducing signalling complex (DISC); caspase-8 and caspase-3 activation | Triggers caspase-mediated apoptosis via death receptor signalling; amplifies cytotoxicity through syncytium formation and immune-mediated tumour cell destruction. | [19, 23, 44] |
| Immunogenic cell death (ICD) | Release of DAMPs (calreticulin, ATP, HMGB1); activation of dendritic cells and cytotoxic T cells via antigen presentation | Stimulates immune recognition of tumour cells, leading to T-cell-mediated destruction; enhances systemic antitumour immunity through cytokine release and tumour antigen exposure, promoting tumour regression. | [24, 39, 68] |
| Viral fusion with cell membranes | HN protein binding to α2,6/α2,3-linked sialic acid receptors; F protein cleavage (polybasic/monobasic) and fusion peptide insertion; HN-F interaction | Facilitates viral entry and genome delivery into cancer cells; promotes syncytium formation, inducing apoptosis and amplifying viral spread within tumour masses, enhancing oncolytic potency. | [11, 41, 73, 74] |

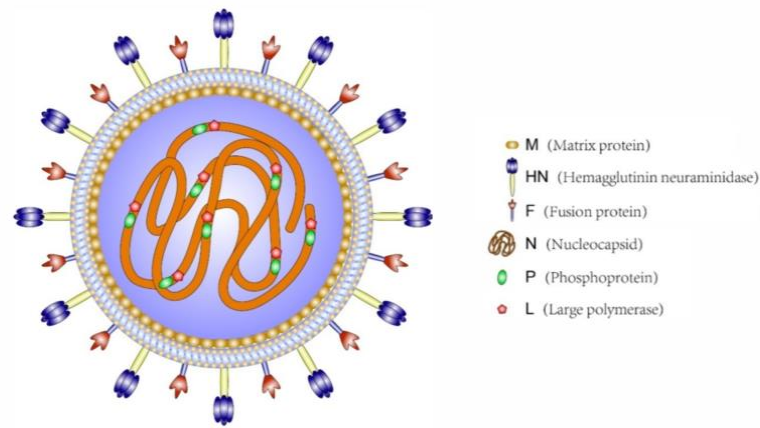

**Figure 1: Morphological and Molecular Structure of NDV virion:** Structure and Composition of *paramyxovirus* type-1. The virion contains nuclear proteins that comprise; NP: Nucleocapsid Protein, P: Phosphoprotein, M: Matrix Protein, F: Fusion Protein, HN: Haemagglutinin–Neuraminidase Protein, L: Large Protein. Among these proteins, NP, P and L combine with the viral RNA to form the ribonucleoprotein complex.

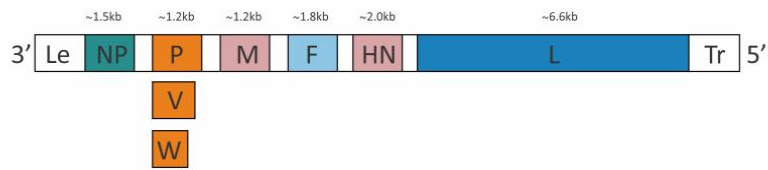

**Figure 2: Schematic diagram of NDV Genome organization:** Genome organization of paramyxovirus type-1 and the minus stranded RNA virus genome encoding genes represented by a bar in 3' to 5' direction.

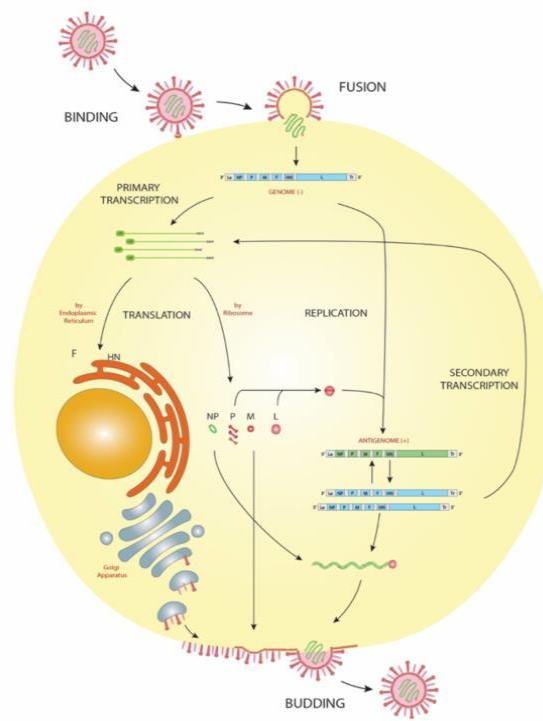

**Figure 3: NDV Infectious Cycle: NDV binds to the surface of the host cell via sialic acid-containing receptor and allows HN to cleave the sialic acid.** Following cleavage of the F protein, the viral genome is released into cytoplasm. The negative sensed viral genome is transcribed into anti-genomic positive stranded mRNA and translated into NP, P, M, F, HN and L proteins. When the NP protein accumulation reaches a certain threshold, the anti-genomic positive stranded mRNA is used to transcribe a full length of negative strand of viral genome. The M, F and HN proteins then move to the membrane for virus assembly of newly produced viral progenies.